\documentclass{stacs_proc}

\usepackage{epsf}
\usepackage{epsfig}
\usepackage{graphicx}
\usepackage{color}
\usepackage{amsmath}
\usepackage{mathrsfs}
\usepackage{amsthm}
\usepackage{latexsym}
\usepackage{amssymb}
\usepackage{amsmath}
\usepackage{algorithm}
\usepackage{algpseudocode}
\usepackage{fancybox}
\usepackage{stmaryrd}
\usepackage{epsf}
\usepackage{verbatim}

\theoremstyle{plain}\newtheorem{claim}[thm]{Claim}
\begin{document}

\title[Truthful Mechanism for Scheduling Unrelated Machines]{An
Improved Randomized Truthful Mechanism for Scheduling Unrelated
Machines}

\author[lab1]{P. Lu}{Pinyan Lu}
\author[lab1]{C. Yu}{Changyuan Yu}
\address[lab1]{Institute for Theoretical Computer Science, Tsinghua University, Beijing, 100084,  P.R. China}  
\email{{lpy, yucy05}@mails.tsinghua.edu.cn}  
\urladdr{http://www.itcs.tsinghua.edu.cn/\{LuPY,YuCY\}}  

\thanks{Supported by the National Natural Science
Foundation of China Grant 60553001
  and the National Basic Research Program of China Grant
  2007CB807900, 2007CB807901. } 

\keywords{truthful mechanism, scheduling} \subjclass{algorithmic
mechanism design}


\begin{abstract}
  \noindent We study the scheduling problem on unrelated machines in the
mechanism design setting. This problem was proposed and studied in
the seminal paper of Nisan and Ronen \cite{NR99}, where they gave a
$1.75$-approximation randomized truthful mechanism for the case of
two machines. We improve this result by a $1.6737$-approximation
randomized truthful mechanism. We also generalize our result to a
$0.8368m$-approximation mechanism for task scheduling with $m$
machines, which improve the previous best upper bound of
$0.875m$\cite{MS07}.
\end{abstract}

\maketitle

\stacsheading{2008}{527-538}{Bordeaux}
\firstpageno{527}

\section{Introduction}\label{intro-sec}
Mechanism design has become an active area of research both in
Computer Science and Game Theory. In the mechanism design setting,
players are selfish and wish to maximize their own utilities. To
deal with the selfishness of the players, a mechanism should both
satisfy some game-theoretical requirements such as
\emph{truthfulness} and some computational properties such as
\emph{good approximation ratio}. The study of their algorithmic
aspect was initiated by Nisan and Ronen in their seminal paper
``Algorithmic Mechanism Design" \cite{NR99}. The focus of that paper
was on the scheduling problem on unrelated machines, for which the
standard mechanism design tools ( VCG mechanisms
\cite{Clarke1971,Groves1973,Vickrey1961})do not suffice. They proved
that no deterministic mechanism can have an approximation ratio
better than $2$ for this problem. This bound is tight for the case
of two machines. However if we allow randomized mechanisms, this
bound can be beaten. In particular they gave a $1.75$-approximation
randomized truthful mechanism for the case of two machines. Since
then, many researchers have studied the scheduling problem on
unrelated machines in mechanism design setting
\cite{JP99,Sourd01,SS02,SX02,GMW07,CKV07,CKK07,MS07}. However their
mechanism remains the best to the best of our knowledge. In a recent
paper \cite{MS07}, Mu'alem and Schapira proved a lower bound of
$1.5$ for this setting. So to explore the exact bound between $1.5$
and $1.75$ is an interesting open problem in this area. In this
paper, we improve the upper bound from $1.75$ to $1.6737$. Formally
we give a $1.6737$-approximation randomized truthful mechanism for
task scheduling with two machines. Using similar techniques of
\cite{MS07}, we also generalize our result to a
$0.8368m$-approximation mechanism for task scheduling with $m$
machines.

Let us describe the problem more carefully. There are $m$ machines
and $n$ tasks, and each machine is controlled by an agent. We use
$t_j^i$ to denote the running time of task $j$ on machine $i$, which
is also called the type value of the agent(machine) $i$ on task $j$.
The objective is to minimize the completion time of the last
assignment (the \emph{makespan}). Unlike in the classical
optimization problem, the scheduling designer does not know $t_j^i$.
Each selfish agent $i$ holds his/her own type values (the $t_j^i$s).
In order to motivate the agents to report their true value $t_j^i$s,
the mechanism needs to pay the agents. So a mechanism consists of an
allocation algorithm and a payment algorithm. A mechanism is called
\emph{truthful} when telling one's true value is among the optimal
strategies for each agent, no matter how other agents behave. Here
the utility of each agent is the payment he/she gets minus the load
of tasks allocated to his/her machine. When randomness is involved,
there are two versions of truthfulness: in the stronger version,
i.e.\emph{ universally truthfulness}, the mechanism remains truthful
even if the agents know the random bits; in the weaker version, i.e.
\emph{truthfulness in expectation}, an agent maximizes his/her
expected utility by telling the true type value. Our mechanisms
proposed in this paper are universally truthful.

Now we can talk about the high level idea of the technical part.
Here we only talk about the allocation algorithms, and the
corresponding payment algorithms, which make the mechanism truthful,
will be given later. First we describe Nisan and Ronen's mechanism
\cite{NR99}. In their mechanism, each task is allocated
independently. For a particular task $j$, if the two values $t^1_j$
 and $t^2_j$ are relatively close to each other, say $t^1_j/t^2_j\in[3/4,4/3]$,
then they allocate task $j$ randomly to machine $1$ or $2$ with
equal probability; if one is much higher then the other, say
$t^1_j/t^2_j>4/3$ or $t^2_j/t^1_j>4/3$, the task $j$ is allocated to
the more efficient machine. The main idea of our mechanism is to
partition the tasks into three categories rather than two. So we
need two threshold values, say $\alpha, \beta$, where
$1<\beta<\alpha$ and a biased probability $r$ , where $1/2<r<1$. If
the two values are relatively close to each other, say
$t^1_j/t^2_j\in[1/\beta,\beta]$, or one is much higher then the
other, say $t^1_j/t^2_j>\alpha$ or $t^2_j/t^1_j>\alpha$ , we do the
same things as Nisan and Ronen's mechanism. In the remaining case,
one is significantly larger than the other, but however still does
not dominate, say $t^1_j/t^2_j\in[\beta,\alpha]$ or
$t^2_j/t^1_j\in[\beta,\alpha]$. In this case, we allocate the task
$j$ to the more efficient one with a higher probability $r$
($r>1/2$) and the less efficient one with a lower probability $1-r$.
The mechanism is quite simple, so it is very computationally
efficient. Intuitively our mechanism will give better approximation
ratios by choosing suitable parameters $\alpha, \beta$ and $r$. This
is indeed true. We can prove an improved approximation ratio of
$1.6737$ by choosing $\alpha=1.4844, \beta=1.1854, r=0.7932$.
However, the proof is quite involved. One reason is that the
situation for the new case (middle case) is
 more complicated than the original two.
The main reason is that their approach becomes infeasible in the
analysis of our mechanism. The proof in Nisan and Ronen's paper is
basically case by case, but unfortunately the number of subcases
increases double exponentially with the number of task types. So we
introduce some substantial new proof techniques to overcome this. We
also think this techniques may further improve the upper bound.

\subsection{Related Work} Scheduling on unrelated machines is one of the
most fundamental scheduling Problems. For this NP-hard optimization
problem, there is a polynomial time algorithm with approximation
ratio of $2$\cite{LST87}. Especially if the number of machines is
bounded by some constant, Angel, Bampis and Kononov gave an
FPTAS\cite{ABK01}. However there is no corresponding payment
strategy to make either of the above allocation algorithms truthful.

The study of this problem in the mechanism design setting is
initiated by Nisan and Ronen. In their paper \cite{NR99}, they gave
a $1.75$-approximation randomized truthful mechanism for two
machines. This result was generalized by Mu'alem and Schapira to a
$0.875m$-approximation randomized mechanism for $m$
machines\cite{MS07}. We improved the two upper bounds to $1.6737$
and $0.8368m$ respectively.

For the lower bound side, Nisan and Ronen gave a lower bound of 2
for deterministic version. This bound was improved by Christodoulou,
Koutsoupias and Vidali to $1+\sqrt{2}$ for $3$ or more machines
\cite{CKV07}. For the randomized version, Mu'alem and Schapira gave
a lower bound of $2-1/m$\cite{MS07}. This also holds for the weaker
notion of truthfulness, i.e., truthfulness in expectation.

Lavi and Swarmy considered a restricted variant, where each task $j$
only has two values of running time , and gave a $3$-approximation
randomized truthful mechanism \cite{LS07}. They first use the cycle
monotonicity in designing mechanisms.

In \cite{CKK07}, Christodoulou, Koutsoupias and Kov{\'a}cs
considered the fractional version of this problem, in which each
task can be split among the machines. For this version, they gave
a lower bound of $2-1/m$ and an upper bound of $(m+1)/2$. We
remark that these two bounds are closed for the case of two
machines as in the integral deterministic version. So to explore
the exact bound for the randomized version seems very interesting
and desirable. We believe that our work in this paper is an
important step toward this objective.

\section{Problem and Definitions}
In this section we review some definitions and results on mechanism
design and scheduling problem.  More details can be found
in\cite{NR99}.

In a mechanism design problem, there are usually some resources to
distribute among $n$ agents. Every agent $i$ has a type value $t^i$,
which denotes his/her preference on the resources. Let
$t=(t^i)_{i\in [m]}$ denote the vector of all agents' type values
and $t^{-i}$ denote the vector of all agents' type vectors except
agent $i$'s. Receiving all the type values $t$ from agents, the
mechanism will produce an output $o(t)=(x(t), p(t))$. Here $x(t)$
specifies the allocation of the resources and is produced by an
allocation algorithm. $p(t)$ specifies the payment to agents and is
produced by an payment algorithm. Every agent $i$ has a valuation
$v^i(x,t^i)$, which describe his/her preference on the output
allocation. The agent $i$'s objective is maximizing his/her utility
function $u^i$, where $u^i=v^i+p^i$, and $p^i$ is the payment
obtained from the mechanism. The mechanism's objective is to
maximize an objective function $g(o,t)$. Formally, we have the
following definitions.

\begin{definition}
A mechanism is a pair of Algorithms $M=(X,P)$.
\begin{itemize}
    \item \emph{Allocation Algorithm $X$:} Its input is $m$ agents' type vectors, $t^1,t^2,\cdots,t^m$, which are reported by the agents.
    and its output is $x=(x^1,\cdots,x^m)$, where $x^i=(x^i_j)_{j\in [n]}\in \{0,1\}^n$ are allocation vector of agent $i$.

    \item \emph{Payment Algorithm $P$:} It outputs a payment vector $p=(p_1,\cdots,p_m)$,
    which depends both on agents' type vectors and allocation vectors produced by allocation algorithm.
\end{itemize}
\end{definition}

A mechanism is \emph{deterministic} if both the allocation algorithm
and payment algorithm are deterministic. When at least one of them
uses random bits, it is called a \emph{randomized mechanism}.

In order to increase utility, an agent may lie when reporting
his/her type values. But for some mechanisms, no agent can increase
his/her utility by lying. This nice property of a mechanism is
called \emph{truthfulness}. We give the formal definitions of
truthfulness.

\begin{definition}
    A deterministic mechanism is \textbf{truthful} iff for every agent, reporting his/her true type values is among
     the best strategies to maximize his/her utility, no matter how other agents
     acts. A randomized mechanism is \textbf{truthful in expectation} iff no
agent can increase his/her expected utility by lying. A randomized
mechanism is \textbf{universally truthful} iff it remains truthful
even if the agents know the random bits.
\end{definition}

From now on, we will only focus on truthful mechanisms. The most
important positive result in mechanism design is generalized
Vickrey-Clarke-Groves(VCG) mechanism
\cite{Vickrey1961,Groves1973,Clarke1971}. Many known truthful
mechanisms are all in VCG family. The mechanisms of VCG family
usually apply to mechanism design problem in which the objective
function is the (weighted) sum of all agents' valuations. To be
formal, we have
\begin{definition}\cite{NR99}
A mechanism $M=(X,P)$ belongs to weighted VCG family if there are real numbers(weights) $\beta^1,\cdots,\beta^n>0$, such that:
\begin{enumerate}
    \item the problem's objective function satisfies $g(o,t)=\sum_i\beta^i v^i(t^i,o).$

    \item $o(t)\in argmax_o(g(o,t)$.

    \item $p^i(t)=\frac{1}{\beta^i}\sum_{i' \neq i}\beta^{i'} v^{i'}(t^{i'},o(t))+h^i(t^{-i})$,
    where $h^i()$ is an arbitrary function of $t^{-i}$.
\end{enumerate}
\end{definition}

\begin{theorem}
(\cite{Roberts1979}) A weighted VCG mechanism is truthful.
\end{theorem}

Now we specify these mechanism notions in the problem of
scheduling unrelated machines.
Assume there are $n$ tasks to be allocated to $m$ machines, each
of which is controlled by an agent. Each agent $i$'s type value is
$t^i=(t^i_j)_{j\in[n]}$, where $t^i_j$ denotes the time to perform
task $j$ on machine $i$.

We use a binary array $x^i=(x^i_j)_{j\in [n]}$ to specify the
allocation of tasks to machine $i$. $x^i_j$ is $1$ if task $j$ is
allocated to machine $i$ and otherwise $0$. Let $x=(x^i)_{i\in
[m]}$ denote the allocation of all the tasks. For an allocation
$x$, agent $i$'s valuation is $v^i=-x^i\cdot t^i$, where $x^i\cdot
t^i=\sum_{j=1}^nx^i_jt^i_j$.

\begin{definition}Given any allocation $x$ of the tasks, the longest running time of the machines is called the
makespan of the allocation. Formally,
makespan$(x)=max_{i\in[m]}x^i\cdot t^i$.
\end{definition}

The objective of the mechanism is to minimize the (expected)
makespan of the allocation. This is not the (weighted) sum of all
agents' valuations. So we can not apply VCG mechanism here. However,
we remark that if there is only one task, the makespan can be viewed
as the sum of all agents' valuations. We will use this observation
in our analysis.

From \cite{NR99} and \cite{MS07}, we know that there is no optimal truthful mechanism
for this problem, even if we allow super-polynomial running time and randomness.
So we will try to find a truthful mechanism with good approximation ratio.

\begin{definition}
Let $t_M(t)$ be the (expected) makespan of the mechanism $M$ on
instances $t$ and $t_{opt}(t)$ be the optimal makespan of instance
$t$. We say mechanism $M$ has approximation ratio $c$ iff for any
instance $t$, $t_M(t)/t_{opt}(t)\leq c$.
\end{definition}

\section{Our Mechanism and the Analysis}
In this  section, we give a truthful scheduling mechanism for $2$
machines case, and show that its approximation ratio is $1.6737$.
Then we generalize our result to the $m$ machines case as in
\cite{MS07} and obtain a $0.8368m$-approximation randomized truthful
mechanism.

\subsection{Generalized Randomly Biased Mechanism}

\begin{center}
\begin{tabular}{|l|}
  \hline
  \textbf{Parameters:} Real numbers $\alpha>\beta\geq 1>r\geq
\frac{1}{2}$ .\\
(Here we choose $\alpha=1.4844,
\beta=1.1854, r=0.7932$.)\\

\textbf{Input:} The reported type vectors $t=(t^1,t^2)$.\\

\textbf{Output:} A randomized allocation $x=(x^1,x^2)$, \\
and a
payment $p=(p^1,p^2)$.\\

\textbf{Allocation and Payment algorithm:}\\

\noindent $x^1_j\leftarrow 0, x^2_j\leftarrow 0,
j=1,2\cdots,n;p^1\leftarrow 0;p^2\leftarrow 0$.\\

For each task $j=1,2\cdots,n$ do\\

\hspace{.1 in} $ s_j \leftarrow \left\{
\begin{aligned}
\alpha, &&\mbox{ with probability } 1-r ,\\
\beta, &&\mbox{ with probability } r-1/2 ,\\
1/\beta, &&\mbox{ with probability } r-1/2 ,\\
1/\alpha, &&\mbox{ with probability } 1-r .\\
\end{aligned}
\right.$\\

\hspace{.1 in} if  $t^1_j<s_j t^2_j$,\\

\hspace{.3 in} $x^1_j=1, p^1\leftarrow p^1+s_j  t^2_j$;\\

\hspace{.1 in} else\\

\hspace{.3 in} $x^2_j=1, p^2\leftarrow
    p^2+s_j^{-1}t^1_j$.\\
\hline
\end{tabular}


\end{center}


\begin{theorem}The Generalized Randomly Biased Mechanism (GBM for short) is
universally  truthful and can achieve a $1.6737$-approximation solution for task
scheduling with two machines.\end{theorem}

We will prove this theorem in the following two subsections. In
\ref{subsec:truthful}, we will prove that our mechanism is
universally truthful. Then we analyze its approximation ratio in
\ref{subsec:approx}

\subsection{Truthfulness}\label{subsec:truthful}

\begin{lemma}The Generalized Randomly Biased Mechanism is
universally truthful.\end{lemma}

\proof To prove that the GBM is universally truthful, we only need
to prove that it is truthful when the random sequence ${s_j}$ is
fixed. Since the utility of an agent equals the sum of the utilities
obtained from each task and our mechanism is task-independent, we
only need consider the case of one task. In this case, say $s_j$ is
fixed and there is only one task $j$, the mechanism is exactly the
VCG mechanism with weight $(1,s_j)$. Since a weighted VCG is
truthful, the GBM mechanism is universally truthful.

\subsection{Estimation of the Approximation Ratio}\label{subsec:approx}
If this subsection, we will estimate the approximation ratio of our
GBM mechanism. Since we already proved that GBM is universally
truthful in \ref{subsec:truthful}, we only need to focus on the
allocation algorithms of GBM. So we can restate the allocation
algorithms for GBM in an equivalent but more understandable way.
Intuitively we should assign one task with larger probability to the
machine which has smaller type value(running time) on it. The idea
of our mechanism is to partition all the tasks into several types
according to the ratio of two agents' type values. For different
types of tasks, we use different biased probabilities to allocate
them. To be formal, we have the following definition.

\begin{definition}
For a task $j$, we call it an $h$-task iff $\frac{t^i_j}{t^{3-i}_j}>
\alpha$ for some $i\in \{1,2\}$; we called it an $m$-task iff
$\beta<\frac{t^i_j}{t^{3-i}_j}\leq \alpha$ for some $i\in \{1,2\}$;
we call it an $l$-task if $\frac{t^i_j}{t^{3-i}_j}\leq \beta$ for
any $i\in \{1,2\}$.\end{definition}

Then, we have the following claim.

\begin{claim} The GBM mechanism
allocates the tasks in the same way as the following allocating
algorithm does.
\begin{itemize}
    \item For $h$-task, we allocate it to the machine with lower
    type value.
    \item For $m$-task, we allocate it to the more efficient machine
    with probability $r$ and to the less efficient machine with probability
    $1-r$.
    \item For $l$-task, we allocate it to two machines with equal
    probabilities.
\end{itemize}
\end{claim}

\proof For each task $j$, we consider the probability that it is
allocated to machine $1$ in GBM. According to the ratio of
$\frac{t^1_j}{t^2_j}$, we have the following $5$ cases:
\begin{itemize}
    \item \emph{Case 1: }¡¡$t^1_j\geq \alpha t^2_j$, then $Pr(x^1_j=1)=0$
    \item \emph{Case 2: }¡¡$\beta \leq t^1_j<\alpha t^2_j$, then $Pr(x^1_j=1)=1-r$
    \item \emph{Case 3: }¡¡$\beta^{-1} \leq t^1_j<\beta t^2_j$, then $Pr(x^1_j=1)=(1-r)+(r-\frac{1}{2})=\frac{1}{2}$
    \item \emph{Case 4: }¡¡$\alpha^{-1}\leq t^1_j<\beta^{-1} t^2_j$, then $Pr(x^1_j=1)=(1-r)+(r-\frac{1}{2})+(r-\frac{1}{2})=r$
    \item \emph{Case 5: }¡¡$t^2_j<\alpha^{-1} t^2_j$, then $Pr(x^1_j=1)=(1-r)+(r-\frac{1}{2})+(r-\frac{1}{2})+(1-r)=1$
\end{itemize}

The probabilities that task $j$ is assigned to machine $1$ by two
algorithms are always the same, so the lemma is true. \qed

\begin{remark}
This claim only says that the (distribution of) allocation produced
by the two methods are the same. However if we use this allocation
algorithm stated in the claim, we can only make the mechanism
truthful in expectation.
\end{remark}

As in \cite{NR99},  we obtain the following crucial claim, which can
help us cut the number of tasks. The proof of this claim is similar
, and we put it in the Appendix.
\begin{claim}\label{claim 3.6}
To analyze the performance of the generalized randomized biased
mechanism, we only need consider the following cases:
\begin{enumerate}
    \item For each $h$-task $j$, the ratio of the two machines' type value is arbitrarily close to $\alpha$. So we can assume it equals $\alpha$.

    \item If $OPT$ allocates an $l$-task $j$ to machine $i$, then $t^{3-i}_j/t^i_j=\beta$.

    \item If $OPT$ allocates an $m$-task $j$ to machine $i$ which has smaller type value, then $t^{3-i}_j/t^i_j=\alpha$.

    \item If $OPT$ allocates an $m$-task $j$ to machine $i$ which has bigger type value, then $t^{3-i}_j/t^i_j=\beta^{-1}$.

    \item One of the machines is more efficient than the other on all $h$-tasks. We assume it's machine $1$.

    \item There are at most $8$ tasks $A,B,C,D,E,F,G,H$. In $OPT$, tasks $A,C,E,G$ are allocated to machine $1$, and the others to machine $2$.
	  Tasks $A, B$ are $h$-tasks. Tasks $C,D,E,F$ are $m$-tasks and tasks $G,H$ are $l$-tasks.
\end{enumerate}
\end{claim}

From the above analysis, we know that we only need to consider the
reduced case as described in Figure \ref{Figure:8task}.

\begin{figure}[htbp]
\begin{center}
\begin{tabular}{|c|c|c|c|c|c|c|}
  \hline
    type & task  & $t_j^1$ & $t_j^2$ & opt-alloc & gbm-alloc(probability) \\
  \hline
  $h_1$ & $A$ & $a$ & $\alpha a $ & 1 & $1:0$ \\
  \hline
  $h_2$ & $B$ &  $b$ & $\alpha b $ & 2 & $1:0$ \\
  \hline
  $m^1_1$ &  $C$ &$c$ & $\alpha c $ & 1 & $r:(1-r)$ \\
  \hline
  $m^1_2$ & $D$ & $d$ & $\beta d $ & 2 & $r:(1-r)$ \\
  \hline
  $m^2_1$ &  $E$ &$\beta e$ & $e $ & 1 & $(1-r):r$ \\
  \hline
  $m^2_2$ &  $F$ &$\alpha f$ & $f $ & 2 & $(1-r):r$ \\
 \hline
  $l_1$ &  $G$ &$g$ & $\beta g $ & 1 & $\frac{1}{2}:\frac{1}{2}$ \\
   \hline
  $l_2$ &  $H$ &$\beta h$ & $h $ & 2 & $\frac{1}{2}:\frac{1}{2}$ \\
  \hline
\end{tabular}
\caption{The Reduced Case.}
\label{Figure:8task}
\end{center}
\end{figure}

Now we can estimate the approximation ratio based on this reduced case.

\begin{lemma}\label{lemma:approx}The allocation produced by \textbf{GBM} is a $1.6737$-approximation
solution
for the task scheduling problem with two machines.\end{lemma}

\proof Let $t_{opt}$ be the make-span of an optimal solution and let
$t_{gbm}$ be the expected makespan of allocations produced by GBM.
We want to show that $t_{gbm}\leq 1.6737t_{opt}$.

From the allocation of the optimal solution, we have that
\[t_{opt}=max\{a+c+\beta e +g, \alpha b +\beta d + f +h\}.\]

Now we will estimate the expected makespan of our mechanism
$t_{gbm}$. First we introduce some notation which will be used in
the following analysis. We will treat the same name $X$
($X=A,B,\cdots,H$) as a random variable, which denotes the
assignment of the task $X$. For example, $C=2$ means that our
mechanism assigns the task $C$ to the second machine. Then the last
column in Figure \ref{Figure:8task} can also be viewed as the
distribution of the random variable $X$ ($X=A,B,\cdots,H$). For
example $Pr(C=1)=r$ and $Pr(C=2)=1-r$. Since our mechanism assigns
each task independently, the random variables are also independent
of each other. More precisely, for any $X,Y\in \{A, B,\cdots, H\}$,
$i,j\in \{1,2\}$ and $X\neq Y$, we have
\[Pr(X=i,Y=j)=Pr(X=i)Pr(Y=j).\]

We use a random variable $M$ to denote the machine finishing last.
More precisely, $M=1$ means the completion time of the first machine
is not earlier than the second machine, otherwise we have $M=2$.

Now we compute the contribution of each task to $t_{gbm}$. Let the
$j$-th task be $X$. Then its contribution to $t_{gbm}$ contains two
parts. First part is from $t_j^1$. $t_j^1$ contributes to $t_{gbm}$
iff our mechanism assigns task $X$ to 1 (e.t. $X=1$) and the machine
1 finishes later (e.t. $M=1$). The situation for $t_j^2$ is similar.
To sum up, the contribution of the $j$-th task $X$ to $t_{gbm}$ is
\[Pr(M=1,X=1)t_j^1+Pr(M=2,X=2)t_j^2.\]

For example, the contribution of task $C$ to $t_{gbm}$ is
\[ Pr(M=1,C=1)c+\alpha Pr(M=2,C=2)c=(Pr(M=1,C=1)+\alpha Pr(M=2,C=2))c.\]

Similarly, we can compute the contribution of each task to
$t_{gbm}$ easily. To simplify the notation, we use $C_x$
($x=a,b,\cdots, h$) to denote the coefficient of $x$ in $t_{gbm}$.
So we have
\[t_{gbm}=C_a a+C_b b+ C_c c+C_d d+C_e e+C_f f+ C_g g+C_h h.\]
where
\begin{eqnarray*}
C_a&=&Pr(M=1),\\
C_b&=&Pr(M=1),\\
C_c&=&Pr(M=1,C=1)+\alpha Pr(M=2,C=2),\\
C_d&=&Pr(M=1,D=1)+\beta Pr(M=2,D=2),\\
C_e&=&\beta Pr(M=1,E=1)+ Pr(M=2,E=2),\\
C_f&=&\alpha Pr(M=1,F=1)+ Pr(M=2,F=2),\\
C_g&=&Pr(M=1,G=1)+\beta Pr(M=2,G=2),\\
C_h&=&\beta Pr(M=1,H=1)+ Pr(M=2,H=2).
\end{eqnarray*}

Since
\begin{eqnarray*}
t_{gbm}&=&C_a a+C_b b+ C_c c+C_d d+C_e e+C_f f+ C_g g+C_h h\\
&=& (C_a a+ C_c c+ \frac{C_e}{\beta} \beta  e+ C_g g) +( \frac{C_b}{\alpha} \alpha b+\frac{C_d}{\beta} \beta d+C_f f+C_h h)\\
&\leq& max(C_a, C_c, \frac{C_e}{\beta}, C_g )(a+c+\beta e +g)+max(\frac{C_b}{\alpha}, \frac{C_d}{\beta}, C_f, C_h)(\alpha b +\beta d + f +h)\\
&\leq& max(C_a, C_c, \frac{C_e}{\beta}, C_g )t_{opt}+max(\frac{C_b}{\alpha}, \frac{C_d}{\beta}, C_f, C_h)t_{opt}
\end{eqnarray*}

So the performance of our mechanism
is bounded by
\[max(C_a, C_c, \frac{C_e}{\beta}, C_g )+max(\frac{C_b}{\alpha}, \frac{C_d}{\beta}, C_f, C_h).\]

We will give bound for every possible sum between $\{C_a, C_c,
\frac{C_e}{\beta}, C_g\}$ and $\{\frac{C_b}{\alpha},
\frac{C_d}{\beta}, C_f, C_h\}$.

First
\begin{eqnarray*}
C_a+\frac{C_b}{\alpha} =Pr(M=1)+\frac{Pr(M=1)}{\alpha} \leq
1+\frac{1}{\alpha}.
\end{eqnarray*}
So $C_a+\frac{C_b}{\alpha}$ is bounded by $1+\frac{1}{\alpha}$.
Later we will choose suitable parameter $\alpha$ so that this value
is not too big.

Now we analyze a more complicated case, say $C_c+ C_f$. Substituting
$C_c$ and $C_f$, we have
\[C_c+ C_f =Pr(M=1,C=1)+\alpha Pr(M=2,C=2)+ \alpha Pr(M=1,F=1)+ Pr(M=2,F=2).\]
Here we use $P_{ijk}$ to denote the joint distribution of three
random variables $M,C,F$ (e.t. $P_{ijk}=Pr(M=i,C=j,F=k)$ ). Then we
can rewrite the formula as following
\[P_{111}+P_{112}+\alpha (P_{221}+P_{222}) + \alpha(P_{111}+P_{121})+(P_{212}+P_{222}).\]
Then we recombine the terms as following
\[(P_{111}+P_{112}+P_{212})+ \alpha (P_{111}+P_{121}+P_{221})+ (1+\alpha) P_{222}.\]
The first term is bounded by $Pr(C=1)$ (since
$Pr(C=1)=P_{111}+P_{112}+P_{212}+P_{211}$); similarly the second
term is bounded by $\alpha Pr(F=1)$; the third term is bounded by
$(1+\alpha) Pr(C=2,F=2)$. So we can bound  $C_c+ C_f$ by
\[Pr(C=1) + \alpha Pr(F=1) +  (1+\alpha) Pr(C=2,F=2)\]
\[= r + \alpha (1-r) + (1+\alpha)r(1-r)=2r+\alpha-r^2-\alpha r^2.\]

Similarly we can bound the remaining $14$ sums as follows. Some of
proofs are slightly more complicated but all of them are along
similar lines. We only list the bounds here, and the details are
omitted here due to the space limitation.

\begin{enumerate}
    \item $C_a+ \frac{C_d}{\beta} \leq 1+\frac{r}{\beta}.$

    \item $C_a+ C_f\leq 1+(1-r)\alpha.$

    \item $C_a+ C_h\leq 1+\frac{\beta}{2}.$

    \item $C_c+\frac{C_b}{\alpha} \leq 1+\frac{1}{\alpha}.$ (Here we use the assumption that $\alpha \leq
1+\frac{1}{\alpha}$.)

    \item $C_c+ \frac{C_d}{\beta}\leq
\frac{r^2}{\beta}+1+r^2+\alpha-r+\alpha r.$

    \item $C_c+ C_h \leq \frac{1}{2}+\frac{1}{2}r+\alpha-\alpha
r+\frac{1}{2}\beta r.$

    \item $\frac{C_e}{\beta}+\frac{C_b}{\alpha} \leq 1+ \frac{1}{\alpha}.$

    \item $\frac{C_e}{\beta}+\frac{C_d}{\beta} \leq (1-r) + \frac{1}{\beta} r + (1+\frac{1}{\beta})r(1-r)\leq  C_c+ C_f.$

    \item $\frac{C_e}{\beta}+C_f \leq \frac{r^2}{\beta}+1+r^2+\alpha-r+\alpha r.$

    \item $\frac{C_e}{\beta}+C_h \leq 1+\frac{r}{2 \beta}+
\frac{1}{2}\beta- \frac{1}{2}r.$

    \item $C_g+\frac{C_b}{\alpha} \leq  1+\frac{1}{\alpha}.$ (Here we use the assumption that $\alpha \leq
1+\frac{1}{\alpha}$.)

    \item $C_g +  \frac{C_d}{\beta} \leq 1+\frac{r}{2 \beta}+
\frac{1}{2}\beta- \frac{1}{2}r.$

    \item $C_g +  C_f \leq \frac{1}{2}+\frac{1}{2}r+\alpha-\alpha
r+\frac{1}{2}\beta r.$

    \item $C_g+C_h  \leq  \frac{3}{4}+\frac{3}{4}\beta.$
\end{enumerate}

To sum up, we have $9$ different bounds: $1+\frac{1}{\alpha}$,
$2r+\alpha-r^2-\alpha r^2$, $1+\frac{r}{\beta}$, $1+(1-r)\alpha$,
$1+\frac{\beta}{2}$, $\frac{r^2}{\beta}+1+r^2+\alpha-r+\alpha r$,
$\frac{1}{2}+\frac{1}{2}r+\alpha-\alpha r+\frac{1}{2}\beta r$,
$1+\frac{r}{2 \beta}+ \frac{1}{2}\beta- \frac{1}{2}r$,
$\frac{3}{4}+\frac{3}{4}\beta$, and one assumption that $\alpha \leq
1+\frac{1}{\alpha}$. We want to choose suitable parameter
$\alpha,\beta,r$ such that the assumption is satisfied and the
maximal bound is as small as possible. This can be easily done
numerically by a mathematical tool such as Matlab. We can choose
$\alpha=1.4844, \beta=1.1854, r=0.7932$. Substituting these values,
we can verify that all the bounds are less than $1.6737$. So we
proved that our mechanism has an approximate ratio of $1.6737$. \qed

\subsection{An Improved Mechanism for $m$ Machines}
As an application of our main result, we turn to the case of $m$
machines. In \cite{NR99}, Nisan and Ronen gave a truthful
deterministic mechanism that achieves an $m$-approximation.
Recently, Mu'alem and Schapira \cite{MS07} generalized Nisan and
Ronen's truthful randomized mechanism for $2$ machines to the case
of $m$ machines. They partitioned the $m$ machines into two sets of
machines with equal size, $S_1$ and $S_2$. Then they construct a new
instance with only two machines, with type values $t_j^i=min_{a\in
S_i}t_j^a, i=1,2$. Applying the mechanism for $2$ machines case,
They showed a universally truthful randomized mechanism that obtains
an approximation of $0.875m$. Using this idea and our improved
result for two machines case, we can improve the ratio from $0.875m$
to $0.8368m$. To be self contained, we give the formal description
of the mechanism here. The proof is similar with \cite{MS07} and
omitted here.

\begin{center}
\Ovalbox{
\parbox[c]{5.8in}{
\textbf{Parameters:} real numbers $\alpha>\beta\geq 1>r\geq
\frac{1}{2}$.

     \textbf{Input:} the reported type value vectors $t=(t^1,t^2,\cdots,t^m)$.

     \textbf{Output:} an randomized allocation
$x=(x^1,x^2,\cdots,x^m)$ and a payment $p=(p^1,p^2,\cdots,p^m)$.

     \textbf{Mechanism:}
\begin{enumerate}
    \item For each machine $i$, let $x^i\leftarrow \emptyset;p^i\leftarrow 0$.
    \item Partition the set of machines into two sets $S_1,S_2$ with equal size. If $m$ is not even, we can add
    an extra machine with infinite type values on every task.
    \item For each task $j$,
    Let $t^a=min_{i\in S_1}t^i_j$, $a=argmin_{i\in S_1}t^i_j$, $t^{a'}=min_{i\in S_1-\{a\}}t^i_j$.
    Let $t^b=min_{i\in S_2}t^i_j$, $b=argmin_{i\in S_1}t^i_j$, $t^{b'}=min_{i\in S_1-\{a\}}t^i_j$.
    \item Apply our mechanism GBM for two machines case to machine $a$ and $b$ on task $j$. Also the payment
    strategy need a little change. If $a$ gets the task, and it will gain a payment $p^a_j$ in GBM, then we pay
    it $min\{p^a_j, t^{a'}_j\}$. If $b$ gets the task, and it will gain a payment $p^b_j$ in GBM, then we pay
    it $min\{p^b_j, t^{b'}_j\}$. This change is in order to keep the mechanism truthful.
\end{enumerate}
 }}
\end{center}

\begin{theorem}
   m-GBM is an universally truthful randomized mechanism for the scheduling problem that obtains an
   approximation ratio of $0.8369m$ when choosing $\alpha=1.4844, \beta=1.1854, r=0.7932$.
\end{theorem}

\section{Conclusions and Open Problems}
This is the first improvement since Nisan and Ronen proposed the
problem and the 1.75-mechanism. We believe it is possible to further
improve the upper bound using our technics. A direct open problem is
to close the gap between the lower bound of $1.5$ and our new upper
bound of $1.6737$.

Another more important direction is to generalize the mechanisms for
2 machine
 to mechanisms for $m$ machines in a more clever way.
 In the general case,
the gap between the best lower bounds (constants) and the best
 upper bounds ($\Theta(m)$) is huge both in deterministic
 and randomized versions. Any improvement in either direction
 is highly desirable.


 \section*{Appendix}
\vskip-0.3cm
\subsection*{Proof of Claim \ref{claim 3.6}}
 \noindent (1) For $h$-task $j$, assume $t^1_j<t^2_j$. We can
 decrease $t^2_j$ to $\alpha t^1_j$, then $t_{gbm}$ will not change
 since GBM always allocates task $j$ to agent $1$. But this may help
 $OPT$, so the approximation ratio can only be worse.

 \noindent (2) Increasing $t^{3-i}_j$ to $\beta t^i_j$ will not
 affect $OPT$ but will increase $t_{gbm}$. This is because the
 probability to allocate $j$ does not change as long as it is still
 an $l$-task, and one type value is increased.

 \noindent (3) It is similar with the above. We can keep increasing
 $t^{3-i}_j$ while $j$ is still $m$-task. Here $\beta \leq
 t^{3-i}_j/t^i_j \leq \alpha $, so we can make it equal $\alpha$.

 \noindent (4) Here $\beta^{-1} \geq t^{3-i}_j/t^i_j \geq \alpha^{-1}
 $, so we can increase $t^{3-i}_j$ until this ratio equals
 $\beta^{-1}$.

 \noindent (5) This is the same as in \cite{NR99}. We omit the proof
 here.

 \noindent (6) Let $h_a,l_a,a\in\{1,2\}$ denote an $h$-task or
 $l$-task respectively which is allocated to agent $a$ in $OPT$. Let
 $m^a_b, a,b\in\{1,2\}$ denote an $m$-task allocated to agent $b$ in
 $OPT$, on which agent $a$ has smaller type value. So there are $8$
 types of tasks. We will prove that any two task $j_1,j_2$ of the
 same type can be combined into a single task $j$ of the same type.
 Firstly, notice that $j_1,j_2$ have the same ratio of the two
 agents' type values. so task $j$ still has this ratio, hence the
 same type. Further more, they are all allocated by GBM with the same
 probability distribution.

 In one direction, combining will leave $t_{opt}$ unchanged.
 Obviously, combining can only increase $t_{opt}$ because any
 allocation obtained for the new instance can be get for the old
 one. Also $t_{opt}$ can be achieved for the new instance since two
 tasks of the same type are allocated to the same agent.

 In the other direction, combining can only increase $t_{gbm}$.\\
 For the $h$-task case, $t_{gbm}$ is also unchanged because GBM always allocate the $h$-tasks to the more efficient agent.\\

 For the $m$-task case, assume $j_1,j_2$ are both
 $m^a_b,a,b\in\{1,2\}$. Let $Y$ denote an allocation of all the
 tasks except task $j_1,j_2$. Let $t_{Y,j_1,j_2}$ (resp.$t_{Y,j}$)
 denote the expected make-span when $j_1,j_2$ (resp. $j$) are (is)
 allocated by GBM and all other tasks are allocated according to
 $Y$. We have to show that $t_{Y,j_1,j_2}\leq t_{Y,j}$.
 Let $T^1,T^2$ denote finishing time of two agents respectively when allocation is $Y$.\\
 If agent $i$ finishes last regardless of how $j_1,j_2$ are
 allocated, then
 \[t_{Y,j_1,j_2}=T^i+r_i(t^i_{j_1}+t^i_{j_2})=t_{Y,j}\] Here $r_i$
 denotes the probability that $j_1,j_2$ and $j$ are allocated to
 agent $i$. Otherwise, if agent $i$ finishes last iff both
 $j_1,j_2$ are allocated to it, then $T^{3-i}\leq
 T^i+t^i_{j_1}+t^i_{j_2}$
 \begin{align*}
 &t_{Y,j_1,j_2}\\
 &=r_i^2(T^i+t^i_{j_1}+t^i_{j_2})+r_i(1-r_i)(T^{3-i}+t^{3-i}_{j_1}+T^{3-i}+t^{3-i}_{j_2})
 +(1-r_i)^2(T^{3-i}+t^{3-i}_{j_1}+t^{3-i}_{j_2})\\
 &\leq (r_i^2+r_i(1-r_i))(T^i+t^i_{j_1}+t^i_{j_2})+((1-r_i)^2+r_i(1-r_i))(T^{3-i}+t^{3-i}_{j_1}+t^{3-i}_{j_2}) \\
 &=r_i(T^i+t^i_{j_1}+t^i_{j_2})+(1-r_i)(T^{3-i}+t^{3-i}_{j_1}+t^{3-i}_{j_2})\\
 &=t_{Y,j}
 \end{align*}
 Finally assume that $t^i_{j_1}\geq t^i_{j_2},i=1,2$ and consider
 the last case where the agent to which $j_1$ is allocated finishes
 last. In this case
 \begin{align*}
 &t_{Y,j_1,j_2}\\
 &=r_i^2(T^i+t^i_{j_1}+t^i_{j_2})+r_i(1-r_i)(T^i+t^i_{j_1})\\
 &+r_i(1-r_i)T^{3-i}+t^{3-i}_{j_1})
 +(1-r_i)^2(T^{3-i}+t^{3-i}_{j_1}+t^{3-i}_{j_2})\\
 &\leq (r_i^2+r_i(1-r_i))(T^i+t^i_{j_1}+t^i_{j_2})+((1-r_i)^2+r_i(1-r_i))(T^{3-i}+t^{3-i}_{j_1}+t^{3-i}_{j_2})\\
 &=r_i(T^i+t^i_{j_1}+t^i_{j_2})+(1-r_i)(T^{3-i}+t^{3-i}_{j_1}+t^{3-i}_{j_2})\\
 &=t_{Y,j}
 \end{align*}

 The $l$-task case is similar with $m$-task case, with
 $r_i=\frac{1}{2}$.
 \qed

\end{document}